\documentclass[prd,aps,showpacs,preprintnumbers,amssymb]{revtex4}
\usepackage{graphicx}

\def\e3p{$\eta \rightarrow 3 \pi$}

\begin{document}

\title{%
\hfill{\normalsize\vbox{%
\hbox{\rm SU}
 }}\\
{Neutrinos with velocities greater than c ?}}

\author{Joseph Schechter $^{\it \bf a}$~\footnote[3]{Email:
 schechte@physics.syr.edu}}

\author{M. Naeem Shahid $^{\it \bf a}$~\footnote[4]{Email:
 mnshahid@physics.syr.edu}}

\affiliation{$^ {\bf \it a}$ Department of Physics,
 Syracuse University, Syracuse, NY 13244-1130, USA,}

\date{\today}

\begin{abstract}

A possible explanation of the results of the OPERA experiment is 
presented. Assuming that the usual value of c should be interpreted 
as the velocity of light in dark matter, we call the ``true"
velocity of light in vacuum, $c_t$. Then the OPERA neutrinos can be 
faster than c but slower than $c_t$. We also discuss the relationship
between $c_t$ and neutrino masses.

\end{abstract}


\maketitle

\section{Introduction}

    The OPERA collaboration \cite{opera} has recently provided evidence 
    that the neutrinos in a beam sent from the CERN to 
    Gran Sasso Laboratories actually traveled faster than the 
    speed of light, $c$. The earlier MINOS Collaboration \cite{minos}
    had found a similar result 
     but with poorer statistics.

     Conventionally the speed of light in vacuum
    is taken \cite{RPP} to be:
    
    \begin{equation} 
        c = 2. 997 924 58 \times 10^8 m/s .
              \label{conventionalc}
              \end{equation}

     In the present note we discuss a simple way in which 
     such a situation can be realized without violating 
     Einstein's Relativity theory (In that theory \cite{bas}, of course,
     the energy of a massive particle with velocity greater than $c$
     would be pure imaginary and hence unphysical).
     
     Our proposed solution first requires the existence \cite{dmpapers} of 
     ``dark matter". The need for such matter is well established 
     both in cosmology \cite{cos} and for the explanation \cite{grc} of the galactic rotation
      curves. There are many proposed dark matter candidates. Here 
      we just assume that the Earth itself is also ``bathed" in dark matter. It 
      is expected that a low strength interaction of dark matter with 
      photons (and perhaps also neutrinos) exists at some level. If these circumstances 
      are granted, all experimental measurements of the velocity of light 
      on and near Earth have presumably {\it not} been carried out for the velocity of 
      light in vacuum but for the velocity of light in dark matter.
      Hence we introduce a dark matter index of refraction $n_{\gamma}$ which is greater
      than unity
      such that,
      
      \begin{equation}
           c= c_{t}/n_{\gamma}
            \label{ctrue}
          \end{equation}
       
       where  $c_t$ is the true velocity of light in vacuum. It is convenient 
       to write,
       
       \begin{equation}
        c_t = c + \Delta, \quad  \Delta > 0.
       \label{Delta}
        \end{equation}   
        
        Clearly $c_t$ should be used instead of Eq. (\ref{conventionalc}) 
        in the formulation of relativistic mechanics.                  
                       
     Of course, since neutrinos would be slowed down, on
     the traversal of the CERN to Gran Sasso path, by ordinary matter, 
     it is necessary to argue that
      neutrinos are slowed down less by interactions in ordinary matter 
      than photons are slowed down in dark matter. The index of refraction for neutrinos 
      in ordinary matter, $n_\nu$ defined by the neutrino velocity $v$
      should be written,
      
      \begin{equation}
       v = c_t/n_\nu
       \label{nuindex}
       \end{equation}
       
       and, for our argument,  should result in $v$ greater than $c$ (or $n_\nu$ 
        less than $n_\gamma$).
        Here we are temporarily
       neglecting, for simplicity, the effect of the small neutrino mass
       in slowing the neutrino. 
       It is convenient to describe the desired requirement by 
       
       \begin{equation}
       v = c + \delta, \quad \delta > 0 .
       \label{veq}
       \end{equation}
       
       Evidently we must have,
       
       \begin{equation} 
      \Delta > \delta.
      \label{Dd}
      \end{equation}
      
      To compare neutrino and photon slowing due to interactions 
      we mention the standard formula \cite{w} for the index of refraction 
      for a wave in a medium:
      
      \begin{equation}
      n = 1 + \frac{2\pi N f(0)}{k^2},
      \label{nformula}
      \end{equation}
      
      where $N$ is the number of scattering centers per unit 
      volume, $k$ is the spatial momentum parameter of the 
      photon or neutrino and $f(0)$ is the forward scattering 
      amplitude for the photon or neutrino scattering off the medium.
      
      The detailed description of either photons or neutrinos interacting
      with dark matter and/or ordinary matter is rather complicated.
      However, at least we may say that the photon case involves
       electromagnetic-strength interaction vertices while the neutrino case
      involves weak-strength interaction vertices. Hence it is certainly plausible 
      that $f(0)_\gamma$ should be considerably larger than
      $f(0)_\nu$. Accepting this, we expect $n_\gamma$ to be considerably
      larger than $n_\nu$.  
     
       As an initial approximation it thus seems reasonable to consider that 
       the neutrino is slowed down just by its nonzero mass while the 
       photon is slowed down by its interaction with dark matter in 
       addition to its interaction with ordinary matter.

 \section{Numbers}
 
 The distance between source and detector for the OPERA experiment
 is listed as:
 
 \begin{equation}
 D = 730.085 \quad km .
 \label{distance}
 \end{equation}
 
 Dividing this by the conventional value of $c$ in 
 Eq. (\ref{conventionalc}) gives a value of the time needed
 for light to traverse $D$ in vacuum:
 
 \begin{equation}
 t_\gamma = 0.002435304454 \quad s
 \end{equation}
 
 The OPERA group determined that the time needed for their $\nu_\mu$
 projectile was,
 
 \begin{equation}
t_\nu = t_\gamma -0.000000058 \quad s =
  0.002435246454 \quad s ,
 \end{equation} 

giving a neutrino velocity,

\begin{equation}
v_\nu = D/t_\nu = 2.997992252 \times 10^8\quad m/s 
\end{equation}

which is larger (at the fifth decimal place) than $c$.
More precisely, including statistical and systematic errors they write:

\begin{equation}
\frac{v_\nu -c}{c} = [2.37 \pm 0.32 (stat)+ 0.34,-0.24 (syst)] \times 10^{-5}.
\end{equation}

Multiplying this equation by $c$ determines the value of $\delta$ defined in 
Eq.(\ref{veq}):

\begin{equation}
 \delta = [7.10 \pm 0.96 (stat) +1.02, -0.72 (syst)] \times 10^3 \quad m/s.
 \label{deltanumber}
 \end{equation}

The value of $\Delta$, or equivalently, the postulated ``true" velocity 
of light $c_t$ is not yet 
determined and will be discussed next.

\section{Neutrino energy}
 
 The OPERA group listed the average neutrino beam energy as
 $E_\nu \approx$ 17 GeV. Introducing an effective neutrino mass,
 $m_{\nu}$, we write the neutrino energy in the usual way:
 
 \begin{equation}
 E_\nu= \frac{m_{\nu}{c_t}^2}{\sqrt{1-{v_{\nu}^2}/{c_t}^2}}.
 \label{nuenergy}
 \end{equation}
 
 Using Eqs (\ref{Delta}) and (\ref{veq}), while
 throwing away the very small terms quadratic in $\Delta$
 and $\delta$, gives for this equation: 
 
 \begin{equation}
 E_\nu \approx \frac{m_{\nu}c^2}{\sqrt{2}} \sqrt{\frac{c}{\Delta-\delta}}
 \label{practicalenergy}
   \end{equation}

Note that $c$ in this equation is the conventional one 
of Eq.(\ref{conventionalc}).
 
Since this formula may look a little unfamiliar, we compare it with 
the analogous formula in standard relativity for the energy of 
a very relativistic particle with velocity $c- \epsilon$ and mass m:

\begin{equation}
E_{standard} \approx \frac{m c^2}{\sqrt{2}} \sqrt{\frac{c}{\epsilon}}.
\label{standardenergy}
\end{equation}

In the present model, the analog of this equation reads:

\begin{equation}
E_{present} \approx \frac{m c{_t}^2}{\sqrt{2}} \sqrt{\frac{c_t}{\epsilon^\prime}},
\label{presentenergy}
\end{equation}

where $\epsilon^\prime = c_t-v$.

     It is clear that rewriting Eq.(\ref{practicalenergy})
as,

\begin{equation}
  \frac{\Delta-\delta}{c} = \frac{(m_{\nu}c^2)^2}{2E_{\nu}^2},
  \label{findctrue}
\end{equation}

enables us to estimate $\Delta-\delta$ in terms of the known beam energy of 
17 GeV and plausible values of the effective neutrino mass, $m_\nu$. The result
is shown, for a reasonable range of effective neutrino masses in Fig. 1. 
Qualitatively, it seems that $\Delta$ is just a tiny bit bigger than $\delta$.
For example if $m_\nu c^2$ = 0.1 eV, $\Delta - \delta$ is about $5 \times 10^{-15}$ m/s. 

\begin{figure}[htbp]
\centering
\rotatebox{0}
{\includegraphics[]{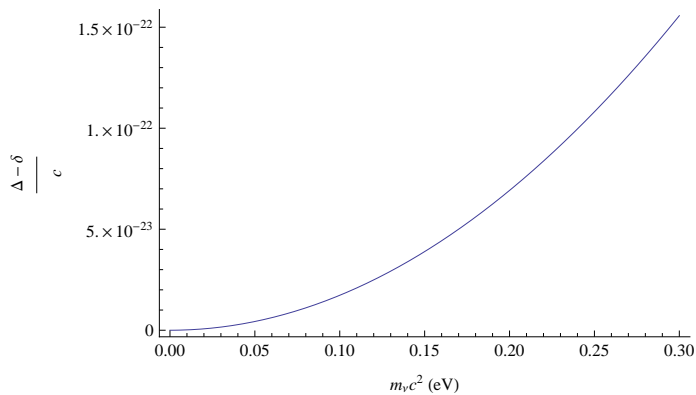}}
\caption[]{Plot of $\frac{\Delta-\delta}{c}$ vs the effective neutrino mass}
\label{triangle}
\end{figure}

\section{Effective neutrino mass}

 In the OPERA experiment, the mu type neutrino $\nu_{\mu}$ was observed. 
 This is, of course, not a mass eigenstate, but a linear combination 
 of the three mass eigenstates $\nu_i$ :
 
 \begin{equation}
 \nu_{\mu} \approx K_{21}\nu_1 + K_{22}\nu_2 + K_{23}\nu_3, 
 \label{masseigenstates}
 \end{equation}
 
 where the $K_{2i}$ are the elements of the leptonic mixing matrix.
 (For this we made the
 approximation that the charged lepton mass matrix in the ``standard model" is 
 diagonal.)
 Thus, some prescription is clearly needed to define the ``average mass of the muon
  type neutrino"
 which was used in the preceding section.
 In models, say for the case of Majorana neutrinos \cite{maj}, the mixing results
 from bringing an ``original mass matrix, $M$ to diagonal form, $\hat{M}$,
 
 \begin{equation}
 \hat{M} = K^{T}MK ,
 \end{equation}

where $K$ is unitary. It is suggestive to identify the average mass of the 
muon neutrino as,  

\begin{equation} 
    m_{\nu} = M_{22} = (K^{-1T})_{2i}\hat{M}_{ii}K^{-1}_{i2}.
\end{equation}

In the approximation where the matrix $K$ is real and using $m_i= (\hat{M})_{ii}$,

\begin{equation}
m_{\nu} \approx (K_{2i})^2 m_i
\end{equation}

 This expression clearly has the correct behavior for the formal limit in 
  which $|K_{22}|$ dominates.
  
  The recent experimental situation concerning the $|K_{2i}|$'s is discussed in 
  \cite{stv}. For definiteness we will adopt the values obtained in the 
  ``tribimaximal" mixing model \cite{HPS}:
  
  \begin{equation}
  K_{21}= 1/\sqrt{6}, \quad K_{22}=1/\sqrt{3}, \quad K_{23}=1/\sqrt{2}.
  \label{tbm}
  \end{equation}
  
 Small corrections to these values have been considered by a number of authors.
  
With the mentioned approximations we get a rough idea of the averaged muon type      
  neutrino mass,
  
  \begin{equation}
  m_\nu \approx m_{1}/6 +m_{2}/3 + m_{3}/2 
   \end{equation}

  At present, the individual neutrino masses, $m_i$ are not known. Only 
  the differences $m_2^2-m_1^2$ and $|m_3^2-m_2^2|$ are available. Thus
  an absolute mass measurement, like one arising from the study of the tritium decay 
  endpoint spectrum is required 
  to complete  the picture. However, we may bound $m_\nu$ by using the bound \cite{cosmobound} from cosmology,
   
    \begin{equation}
 ( m_1 + m_2 + m_3) c^2 < 0.3 \quad {\rm eV}.
  \label{cosmobound}  
   \end{equation}  
 
Introducing the  3-vectors ${\bf A}=(1/6,1/3,1/2)$ and ${\bf m}=(m_1,m_2,m_3)$ 
we write,  
   \begin{equation}
  m_\nu= {\bf A}\cdot {\bf m} < |{\bf A}||{\bf m}|=\frac{\sqrt{14}}{6}|\bf{m}|.
  \label{mnubound}  
   \end{equation}  
 Using the positivity of the $m_i$'s we note, 
   \begin{equation}
|{\bf m}| <m_1 + m_2 + m_3. 
  \label{positivity}  
   \end{equation}  

With the help of Eq. (\ref{cosmobound}) we get   

 \begin{equation}
m_\nu c^2< 0.3 \frac{\sqrt{14}}{6} = 0.187 \quad {\rm eV}. 
  \label{positivity}  
   \end{equation}   
 
 From Fig. (1) we finally read off the bound on the extremely small deviation $\frac{\Delta - \delta}{c}$,
\begin{equation}
\frac{\Delta - \delta}{c} < {\rm 7 \times 10^{-23}}. 
  \label{positivity}  
   \end{equation}   
  
       This very small value seems to be compatible with a bound of about 10$^{-20}$  \cite{bw}
        on non trivial contributions to the dispersion relation (Cauchy formula) for free space.

  \section{Summary and Discussion}
 
     We discussed a scenario which could conceivably explain the apparent 
     observation of neutrinos traveling faster than the conventional velocity of light in
     vacuum. It would require the Earth to be "bathed" in hard to detect dark matter 
     which would nevertheless have an interaction at some level with light. Then experiments 
     measuring $c$ in what was considered to be vacuum would have been measuring the velocity
     of light through dark matter. Presumably the ``true" velocity of light, $c_t$ would be 
     greater than c and neutrinos could violate the $c$- bound as long as they obeyed the $c_t$-
     bound. The "true" velocity of light in this picture turns out to be just an 
     extremely tiny bit larger than the neutrino speed measured by the OPERA collaboration.
  
     In this work we concentrated on explaining the OPERA results and did not 
     discuss the well known fact that neutrinos arrived at earth before light from the 
     supernova 1987A. This is certainly important data to be understood but 
     we felt a full interpretation involves knowing the relative times of emission for 
     the neutrinos and the light. This is apparently not fully established; presumably
     the observation of another supernova will be helpful in this regard. On the other hand, 
     it is very likely that additional accelerator study of neutrino flight times will 
     be soon be carried out at OPERA and elsewhere. 
     
     The present paper also includes (in sections III and IV) an apparently new
      ``effective mass" approach for conveniently treating the motion of the mu-type neutrino
       which is, of course, not a mass eigenstate. 
     
 \section*{Acknowledgments} \vskip -.5cm 
We would like to thank J. Cline for a helpful comment on this manuscript.
 This work was supported in part by the U. S. DOE under Contract no. DE-FG-02-85ER 40231.

\end{document}